\documentclass[a4paper,twocolumn,english]{IEEEtran}
\usepackage[T1]{fontenc}
\usepackage{color}
\usepackage{amsbsy}
\usepackage{amssymb}
\usepackage{graphicx}
\usepackage{amsmath}
\usepackage{multicol,lipsum}
\usepackage[final]{changes}
\colorlet{Changes@Color}{red}
\usepackage[ruled, lined, longend, linesnumbered]{algorithm2e}
\usepackage[noend]{algpseudocode}
\usepackage{mathtools}
\makeatletter
\def\BState{\State\hskip-\ALG@thistlm}
\makeatother

\makeatletter

\pdfpageheight\paperheight
\pdfpagewidth\paperwidth

\usepackage{subfigure}
\usepackage{cite}
\def\authorblockA{\@IEEElegacywarn{authorblockA}{IEEEauthorblockA}\IEEEauthorblockA}
\def\authorblockN{\@IEEElegacywarn{authorblockN}{IEEEauthorblockN}\IEEEauthorblockN}

\makeatother

\usepackage{babel}
\begin{document}

\title{Near-field Ultra-wideband mmWave Channel Characterization Using Successive Cancellation Beamspace UCA Algorithm}

\author{Fengchun Zhang and Wei Fan 

	\thanks{Fengchun Zhang and  Wei Fan are with the APMS section at
		the Department of Electronic Systems, Faculty of Engineering and
		Science, Aalborg University, Denmark (email: \{fz, wfa\}@es.aau.dk. (Corresponding author: Wei Fan)}  } %

\maketitle
\begin{abstract}
	Of the wide palette of 5G features, ultra-wide  bandwidth and large-scale antenna configuration  are regarded as the essential  enabling technology components at millimeter wave (mmWave) communication. Accurate knowledge of delay and angle information of multipath components is essential for many applications in mmWave systems. There is a strong need for a low computation-cost channel estimation algorithm for such systems, where typically adopted far-field and narrowband assumptions might be violated. {In this work, a generic yet novel beamspace uniform circular array (UCA) beamforming algorithm with successive cancellation scheme is proposed to jointly detect the impinging angle and delay of the multipath components. The proposed algorithm is computationally cheap and it works for ultra-wideband (UWB) systems in the near-field conditions.} Both numerical simulations and experimental validation results are provided to demonstrate the effectiveness and robustness of the proposed algorithm, compared to the state-of-art works.  
 \end{abstract}

\begin{IEEEkeywords}
Beamspace UCA, phase mode excitation, uniform circular array, near-field, ultra-wideband, mmWave channel characterization, large-scale antenna systems
\end{IEEEkeywords}

\section{Introduction}

The next generation wireless communication system, typically referred to as the fifth generation (5G), is currently under
intensive research and development. Millimeter wave (mmWave) communication is regarded as one of the enabling components for 5G cellular communication systems, thanks to the large amount of available frequency spectrum \cite{rappaport2013mwave,andrews2014will,demestichas20135g,Ruisi_1, Ruisi_2,Ke_2}. However, mmWave  transmissions suffer from high propagation loss and blockage, unlike sub-6GHz legacy frequency band. High antenna gain (typically realized with a large-scale antenna structure) is required to overcome the radio propagation loss and achieve the good signal-to-noise ratio (SNR) at mmWave bands. Therefore, mmWave transmissions will be potentially ultra-wideband (UWB) (e.g. with an absolute bandwidth larger than 500 MHz) and involve large-scale antenna systems at both communication ends \cite{rusek2013scaling,larsson2014massive}. 

It is of importance to understand how the radio signal propagates in the given scenario. A new air interface system  design typically starts with the understanding of the deployment scenario, where channel models should be developed to reflect the physical transmission.
Accurate knowledge of the delay and angle information of multipath components is essential for many applications in mmWave systems.
\added{The extracted multipath components can be further utilized for analysis, clustering and modeling of wireless channels, which is fundamental for system design and performance evaluation \cite{gustafson2014mm,he2018clustering}. }
A popular channel sounding setup at mmWave bands is to utilize a nonreal-time channel sounder based on the virtual antenna array concept \cite{dohler2002vaa,medbo2015vcua,allan2018vcua,Ke_1}. This strategy is simple, flexible and cost-effective, though it is not suitable for dynamic propagation scenarios due to slow mechanical movement of antennas and channel sounding time for each virtual array location. Furthermore, the mutual coupling effects between the array elements are not presented, which is beneficial since it can further simplify the channel estimation. Virtual uniform circular array (UCA) is particularly popular, since it only requires radial mechanical movement of the positioner and presents approximately constant beam patterns over $360^o$ azimuth angle, regardless of the steering angle \cite{davies1983uca,long1967uca}. 

\replaced{Channel estimation algorithms proposed in the literature for power-angle-delay profile (PADP) estimation were mainly under the far-field and narrowband assumptions. The far-field assumption holds when the distance between the scattering source and the antenna array is larger than the Fraunhofer distance $2D^2/\lambda$, with $D$ the antenna array aperture and $\lambda$ the wavelength. The far field assumption can be violated in short-range mmWave propagation scenarios. The narrowband assumption holds when the condition $D/\lambda \ll f/B$ is satisfied, where $f$ is the frequency and $B$ is the bandwidth. When the system bandwidth becomes wide (e.g. in the mmWave context) that the narrowband assumption does not hold anymore for a given array aperture and frequency, the propagation delay of each path can be resolved at different delay bins between array elements. The existing algorithms in the literature would fail to work under near-field and wideband scenarios, as briefly summarized below.}{} 
\begin{itemize}	
		\item[1)] \added{Beamforming techniques such as classical beamforming (CBF) and Capon beamforming.  As demonstrated in \cite{fan2016eurasip}, The beam pattern of the CBF method is highly sensitive to near-field conditions (which would introduce a power loss in the target direction) and UWB bandwidth (which would introduce a joint sidelobe in the delay and angle domain and difficult to suppress \cite{fan2016eurasip}). Capon beamformer degrades as well in near-field and UWB conditions, since the steering vectors are typically calculated at the center frequency and under far-field conditions.}
		\item[2)] \added{Subspace methods. The conventional subspace angle estimation methods are based on the element-space covariance matrix, where the steering vector is the function of center frequency. Thus, it implied the narrowband assumption. UCA is attractive in angle estimation also thanks to its circular symmetry, where beamspace transform of UCA based on the phase mode excitation principle can be applied \cite{tewfik1992bs,gentile2008fi,chan2002design}. Basically, we can transform the UCA array manifold (in the element space) into uniform linear array (ULA) manifold (in the phase mode space), which enables us to develop computationally efficient and high-resolution subspace algorithms \cite{zoltowski1992direction,zoltowski1996closed,pesavento2002direction,belloni2003unitary,mathews1994_2d}. However, the aforementioned beam-space multiple signal classification (MUSIC) or estimation of singal parameters by rotational invariance techniques (ESPRIT) algorithms were only investigated under narrowband and far-field assumptions. Subspace methods were also developed for joint angle-delay estimation (JADE). With the JADE strategy, the JADE-MUSIC \cite{van1997jade_music} and the JADE-ESPRIT \cite{van1997jade} algorithms exploited both space and time properties by stacking the array element channel impulse responses into a high dimension vector. After performing the eigen-decomposition on the covariance matrix of the constructed vector, JADE can be obtained either with MUSIC or ESPRIT methods. However, both algorithms are unfavorable for the UWB large-scale antenna systems, due to the high complexity introduced by the high dimension of the stacked space-time vector.	}
	\item[3)] High resolution parametric methods. Maximum likelihood estimator (MLE) is a popular high resolution channel estimation algorithm. However, the MLE is well-known for its high computational complexity due to its joint estimation mechanism, particularly when the channel  parameter dimension is large. An attempt to reduce the MLE computation complexity was proposed in \cite{ji2017_mle,ji2018channel}. The space-alternating generalized expectation-maximization (SAGE) algorithm is a relatively low-cost expectation-maximization (EM) algorithm, where the multi-dimensional search is replaced with several one-dimensional searches \cite{bernard1999sage}. However, a prerequisite to utilize the SAGE algorithm is that the likelihood function needs to be independent between different parameter spaces, which might be violated when narrowband and far-field assumptions do not hold, e.g. for the UWB large-scale antenna systems \cite{bernard1999sage}. A general spherical wave model, i.e. including the distance to the scatterer, can be introduced in the MLE type algorithm implementation, which would, however, further increases the computation complexity. 
\end{itemize}

\added{There is a strong need for a low-cost and generic channel estimation algorithm for joint angle and delay profile estimation, which works in practical propagation scenarios, regardless of the antenna system scale (i.e. large or small) and frequency bandwidth settings (i.e. narrow or wide). That is, it can be applied for the near-field UWB scenarios without increasing the algorithm computation complexity. However, such algorithm is missing in the literature, to the best knowledge of the authors. In this work, a novel low-cost beamspace UCA beamforming algorithm with a successive cancellation scheme is proposed to jointly detect the angle and delay of the multipath components. The proposed algorithm is insensitive to the system bandwidth and distance to the scatterers, making it suitable for UWB large-scale antenna systems. In this paper, we firstly demonstrated the performance deterioration of CBF of the UCA in the element space when far-field assumption does not hold. To solve the issue, we resort to the beamspace transform of UCA based on the phase mode excitation principle. The beamspace UCA is shown to be robust to phase errors introduced by the spherical wavefront. A novel beamspace UCA algorithm with the successive cancellation of the detected paths is detailed. Both numerical simulations and experimental results at mmWave bands are provided to demonstrate the effectiveness and robustness of the proposed method in channel parameter estimation for UWB large-scale antenna systems.}

\section{Problem Statement}
\textcolor{black}{Assume a UCA is distributed in the x-y plane and its center is located at the origin of the coordinate system. The UCA consists of $P$ isotropic antenna elements with radius $r$. The angular position of the $p$-th element is $\varphi_{p}=2\pi\cdot(p-1)/P$, $p\in[1,P]$. Suppose there are $N$ paths impinging at the UCA, the channel frequency response at the $p$-th UCA element is the superposition of the channel responses of the $N$ paths,
\vspace*{-.01in}
\begin{equation}
H_p(f)=\sum_{n=1}^{N}  \alpha_n \exp(-j2\pi f \tau_n) \cdot  a_p(f,\mathbf{\Theta}_n),\label{eq:element frequency response}
\end{equation}
where $\alpha_{n}$ and $\tau_{n}$ represent the complex amplitude and delay of the $n$-th path, respectively. $\mathbf{\Theta}_n = [D_n, \theta_n, \phi_n]$ denotes the scatterer location vector of the $n$-th path, where $D_n$ is the distance between the $n$-th scatterer and the UCA center, $\theta_n$  and $\phi_n$ the elevation and azimuth angle of the $n$-th path, respectively. $a_p(f,\mathbf{\Theta}_n)$ is the transfer function between the $n$-th scatterer and the $p$-th UCA element, which is normalized by the transfer function between the $n$-th scatterer and the UCA center, as
\vspace*{-.01in}
\begin{equation}
a_p(f,\mathbf{\Theta}_n)=\beta_{p,n}\exp(-j k \varPsi_{p,n}).\label{eq:a_p}
\end{equation}
 $\beta_{p,n} = D_n/d_{p,n}$ denotes the relative path loss term at the $p$-th element with respect to (w.r.t) the UCA center, where $d_{p,n}$ represents the distance between the $n$-th scatterer and the $p$-th UCA element. Under far-field assumption, we have $\beta_{p,n}=1$.
 $k = 2 \pi f /c$ is the wave number with $c$ denoting the speed of light.
 $\varPsi_{p,n}$ indicates the relative propagation distance to the $p$-th UCA element w.r.t UCA center, i.e.,
\vspace*{-.01in}
\begin{equation}
\varPsi_{p,n} =d_{p,n} - D_n,\label{eq:r_dis}
\end{equation}
where the distance term $d_{p,n}$ is given by
\begin{equation}
d_{p,n}= \sqrt{D_n^2 + r^2 - 2rD_n\sin{\theta_n}\cos(\phi_n-\varphi_p)}.\label{eq:dis}
\end{equation}
 Following Taylor series expansion, we can approximate $d_{p,n}$ as
 \begin{equation}
 d_{p,n}= D_n - r\sin{\theta_n}\cos(\phi_n-\varphi_p) + \varDelta_{p,n},\label{eq:dis_a}
 \end{equation}
 where $\varDelta_{p,n}$ is the term introduced by the near-field condition. When the plane-wave assumption holds, we have $\varDelta_{p,n}=0$.
 By substituting (\ref{eq:dis_a}) into (\ref{eq:r_dis}), $\varPsi_{p,n}$ can be simplified as 
 \begin{equation}
 \varPsi_{p,n}=  - r\sin{\theta_n}\cos(\phi_n-\varphi_p) + \varDelta_{p,n}.\label{eq:r_dis_a}
 \end{equation}}
 
 In this paper, the objective is to  detect path parameters $\{\alpha_n, \tau_n, \phi_n\}$ with arbitrary elevation angle $\theta_n$ present in the near-field condition for $n\in[1,N]$. The performance of CBF and frequency-invariant beamforming (FIBF) based on UCA for a general 3D multipath scenario \added{under far-field assumption} has been discussed in \cite{zhang2017fi}. Below, we limit our discussions on the beamforming properties of CBF and FIBF under near-field conditions, assuming that all impinging paths are confined to the plane of the UCA, i.e. with elevation angle $\theta_n = \frac{\pi}{2}$ for $n\in [1,N]$ for simplicity. Though the proposed algorithm works for arbitrary 3D propagation scenarios as later discussed in section \ref{simulation}  and \ref{measurement}. 


\subsection{CBF in the UCA element space}  \label{cb_ff}
\begin{figure}
	\begin{centering}
		\textsf{\includegraphics[bb=100bp 260bp 695bp 565bp,clip,scale=0.6]{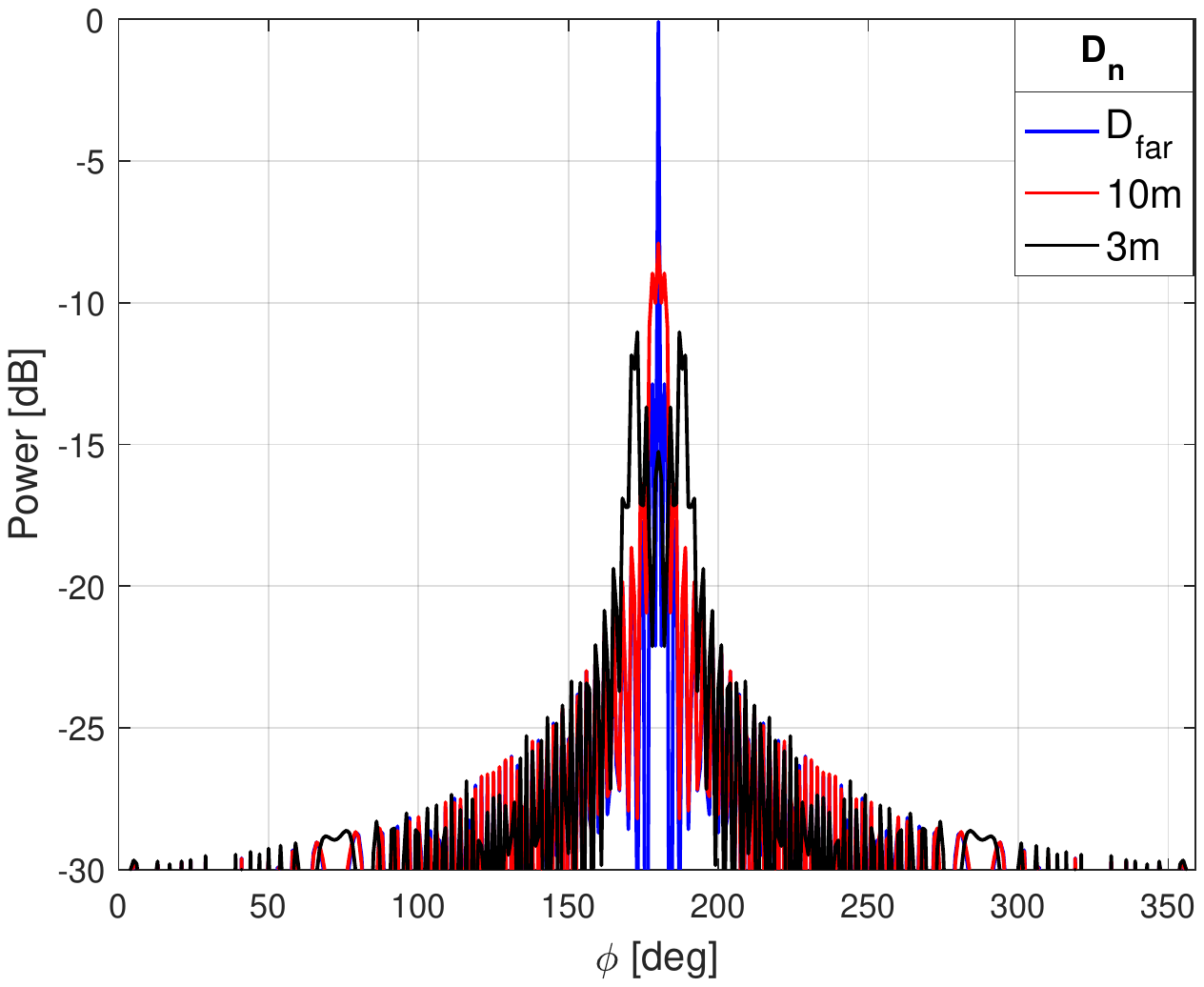}}
		\par\end{centering}
	
	\caption{The unit CBF beam pattern $v_{n}(f,\phi)$ in the UCA element space for different $D_n$ with $\theta_n = \pi/2$, $\phi_n = \pi$, $r$ = 0.5 m, $f$ = 29 GHz and $P=720$ are set for the UCA. The calculated Fraunhofer far-field distance $D_{far} = 2(2r)^2/\lambda$ is around 193 m. }
	\label{bp_cb_dis}
\end{figure}

\textcolor{black}{Using the CBF under plane-wave assumption, the steering weight of the $p$-th UCA element can be written as
\begin{equation}
w_{p}(f,\phi)=\frac{1}{P} \exp[-k r \cos(\phi-\varphi_p)].\label{eq:w_cb}
\end{equation}
Therefore, the UCA array beam pattern can be obtained by taking the coherent summation of the element responses as
\begin{equation}
\begin{split}
B(f,\phi)& = \frac{1}{P}\sum_{p=1}^{P} w_p(f,\phi) \cdot H_p(f).\\        
\end{split}
\label{eq:b}
\end{equation}
By taking (\ref{eq:element frequency response}) into (\ref{eq:b}), we have
\begin{equation}
\begin{split}
B(f,\phi)& = \sum_{n=1}^{N} \alpha_n \exp(-j2\pi f \tau_n)\cdot \sum_{p=1}^{P} w_p(f,\phi) \cdot a_p(f,\mathbf{\Theta}_n)/P   \\   
         & = \sum_{n=1}^{N}  \alpha_n \exp(-j2\pi f \tau_n) \cdot v_n(f,\phi)  \\
         & = \sum_{n=1}^{N} B_n(f, \phi),    
\end{split}
\label{eq:b_1}
\end{equation}
where $B_n(f, \phi)$ represents the beam pattern of the $n$-th path, i.e. the beam pattern of a single path. As seen from (\ref{eq:b_1}), the linear superposition of the CBF beam patterns of all paths gives the beam pattern of the multiple paths.}
\textcolor{black}{$\arrowvert v_n(f,\phi)\arrowvert$ denotes the unit beam pattern term of the $n$-th path with CBF. In ideal case, $\arrowvert v_n(f,\phi)\arrowvert$ mimics a Dirac delta function in $\phi$ domain, i.e. $\arrowvert v_n(f,\phi)\arrowvert  = \delta(\phi-\phi_n)$. Thus the peak location of $\arrowvert  v_n(f,\phi) \arrowvert $ gives the estimate of $\phi_n$, $\alpha_n$ and $\tau_n$ can be obtained via inverse Fourier transform (IFT) of $B(f,\phi)$ at $\phi = \phi_n$.
According to (\ref{eq:b_1}), (\ref{eq:w_cb}) and  (\ref{eq:a_p}), $v_n(f,\phi)$ can be given by
 \begin{equation}
 \begin{split}
 v_{n}(f,\phi)=& \frac{1}{P} \sum_{p=1}^{P}\beta_{p,n} \cdot \exp(-jk \varDelta_{p,n}) \\
               &\cdot \exp\{-jkr[\cos(\phi-\varphi_p)-\cos(\phi_n-\varphi_p)]\}.\label{eq:v_cb}
 \end{split}
 \end{equation}
Under the far-field condition (i.e. $\beta_{p,n}=1$ and $\varDelta_{p,n} = 0$), we have $\arrowvert v_{n}(f,\phi_n)\arrowvert = 1$ and $ \arrowvert v_{n}(f,\phi_{n'})\arrowvert <1$ with $\phi_{n'}\neq \phi_{n}$, as illustrated by the blue line in Fig.\ref{bp_cb_dis}. Thus the peak location in the beam pattern gives the angle estimate of the $n$-th path. }

\added{To investigate whether $\arrowvert v_{n}(f,\phi)\arrowvert$ can still maintain this property under near-field condition, we reduce the distance $D_n$ and its effect on the beam patterns of $v_{n}(f,\phi)$ is shown in Fig. \ref{bp_cb_dis}, i.e. the red and black curves, as examples. The results show that the beam patterns are highly susceptible to phase errors introduced by the spherical wavefront in the near-field condition.
The near-field condition results in the main beam distortion of the beam pattern, including the power loss (the red curve) and the concave pattern in the target direction (the black curve). The power loss results in underestimating the amplitude $\alpha_n$ and the concave pattern would result in fake path angle detection.}
 
\added{To further investigate the power loss in the target direction under near-field condition, the distance from $3$ m to $70$ m was set and the result is shown by the blue dotted line in Fig. \ref{cbf_fibf_dis2}. The figure shows that with CBF method, a power loss more than $35$ dB might exist in the target direction for typical indoor scenarios.}
 
Therefore, besides the joint sidelobe problem introduced by the UWB bandwidth in PADP as shown in \cite{fan2016eurasip}, the CBF of UCA in the element space would also suffer from significant power loss in target directions or even failure of detecting the true paths, due to phase errors introduced in the near-field condition, which makes it unsuitable for channel parameter estimation of mmWave UWB large-scale antenna systems.  

\subsection{Beamformer in the UCA phase mode space}

\textcolor{black}{The frequency response of the UCA element space can be converted to phase mode space as
\begin{equation}
\begin{split}
\underline{H}_m(f) &= \frac{1}{P}\sum_{p=1}^{P}\hat{G}_m(f) \cdot \exp(jm\varphi_p) \cdot H_{p}(f),\\
\end{split}
\label{eq:hm}
\end{equation}
where $\underline{H}_m(f) $ denotes the $m$-th mode response of the UCA in the phase mode space.
As discussed in \cite{zhang2017fi}, $\underline{H}_m(f)$ can be approximated to $\underline{H}_m(f) = \exp(jm\phi_n)$ under far-field condition and with all paths confined in the UCA plane, i.e. $\theta_n = 90^o, n\in[1, N]$, where the compensation filter was defined as $\hat{G}_m(f) = 1/[j^m J_m(kr)]$ with $J_{m}(\cdot)$ denoting the Bessel function of the first kind with
order $m$. However, in practical propagation scenario, it is very unlikely that the incident paths are strictly limited to the UCA plane. Therefore, we modified the compensation filter for 3D propagation scenarios as \cite{zhang2017fi},
\begin{equation}
\frac{1}{\hat{G}_m(f)}= 0.5j^{m}[J_{m}(kr)-jJ_{m}^{'}(kr)],\label{eq:Gm}
\end{equation}
 where 
  $(\cdot)^{'}$ denotes the differential operator.}

\textcolor{black}{By taking equation (\ref{eq:element frequency response}) into (\ref{eq:hm}) and changing the order of the two summations, we have
\begin{equation}
\begin{split}
\underline{H}_m(f) =&  \sum_{n=1}^{N} \alpha_n \exp(-j 2 \pi f \tau_n)\\
                  & \cdot \bigg\{\frac{1}{P}\sum_{p=1}^{P}\hat{G}_m(f) \exp(jm\varphi_p)\cdot a_p(f,\mathbf{\Theta}_n) \bigg\}\\
                  =& \sum_{n=1}^{N} \alpha_n \exp(-j 2 \pi f \tau_n) \cdot \underline{a}_m(f,\mathbf{\Theta}_n),
\end{split}
\label{eq:hm_1}
\end{equation}
where $\underline{a}_m(f,\mathbf{\Theta}_n)$ is given by the $\{\cdot\}$ term in the above equation.
Comparing the above equation with equation (\ref{eq:element frequency response}), we can see that for the $n$-th path, $\underline{a}_m(f,\mathbf{\Theta}_n)$ is the UCA manifold of the $m$-th mode in the phase mode space and $a_p(f,\mathbf{\Theta}_n)$ defined in (\ref{eq:a_p}) is the UCA manifold of the $p$-th element in the element space.  }

\textcolor{black}{The beam pattern of the FIBF can be written as
\begin{equation}
\underline{B}(f,\phi)= \frac{1}{2M+1} \sum_{m=-M}^{M}  \exp(-jm\phi) \cdot \underline{H}_m(f), \label{eq:aoa_fi}
\end{equation}
where $\exp(-jm\phi)/(2M+1)$ is the steering weight of the $m$-th phase mode.
Substituting the array phase mode response $ \underline{H}_m(f)$ defined in (\ref{eq:hm_1}) into the above equation and rearranging the summation order, we have
\begin{equation}
\begin{split}
\underline{B}(f,\phi)&= \sum_{n=1}^{N}  \alpha_n \exp(-j 2 \pi f \tau_n) \cdot \underline{v}_n(f,\phi)\\
                     &= \sum_{n=1}^{N} \underline{B}_n(f,\phi), \label{eq:aoa_fi_1}
\end{split}
\end{equation}
where $\underline{B}_n(f,\phi)$ represents the FIBF beam pattern of the $n$-th path, i.e. the FIBF beam pattern of a single path. The FIBF beam pattern of the multiple paths is obtained by the linear superposition of the FIBF beam patterns of all paths. Similar to the unit beam pattern term $v_n(f,\phi)$ of CBF,  $\underline{v}_n(f,\phi)$ indicates the unit FIBF beam pattern term of the $n$-th path, which is expressed by
\begin{equation}
\underline{v}(f,\phi)= \frac{1}{2M+1} \sum_{m=-M}^{M} \underline{a}_m(f,\mathbf{\Theta}_n) \exp(-jm\phi), \label{eq:v_fi}
\end{equation}
 where $\underline{a}_m(f,\mathbf{\Theta}_n)$ is given in (\ref{eq:hm_1}).}
 
\begin{figure}
	\begin{centering}
		\textsf{\includegraphics[bb=100bp 260bp 695bp 565bp,clip,scale=0.6]{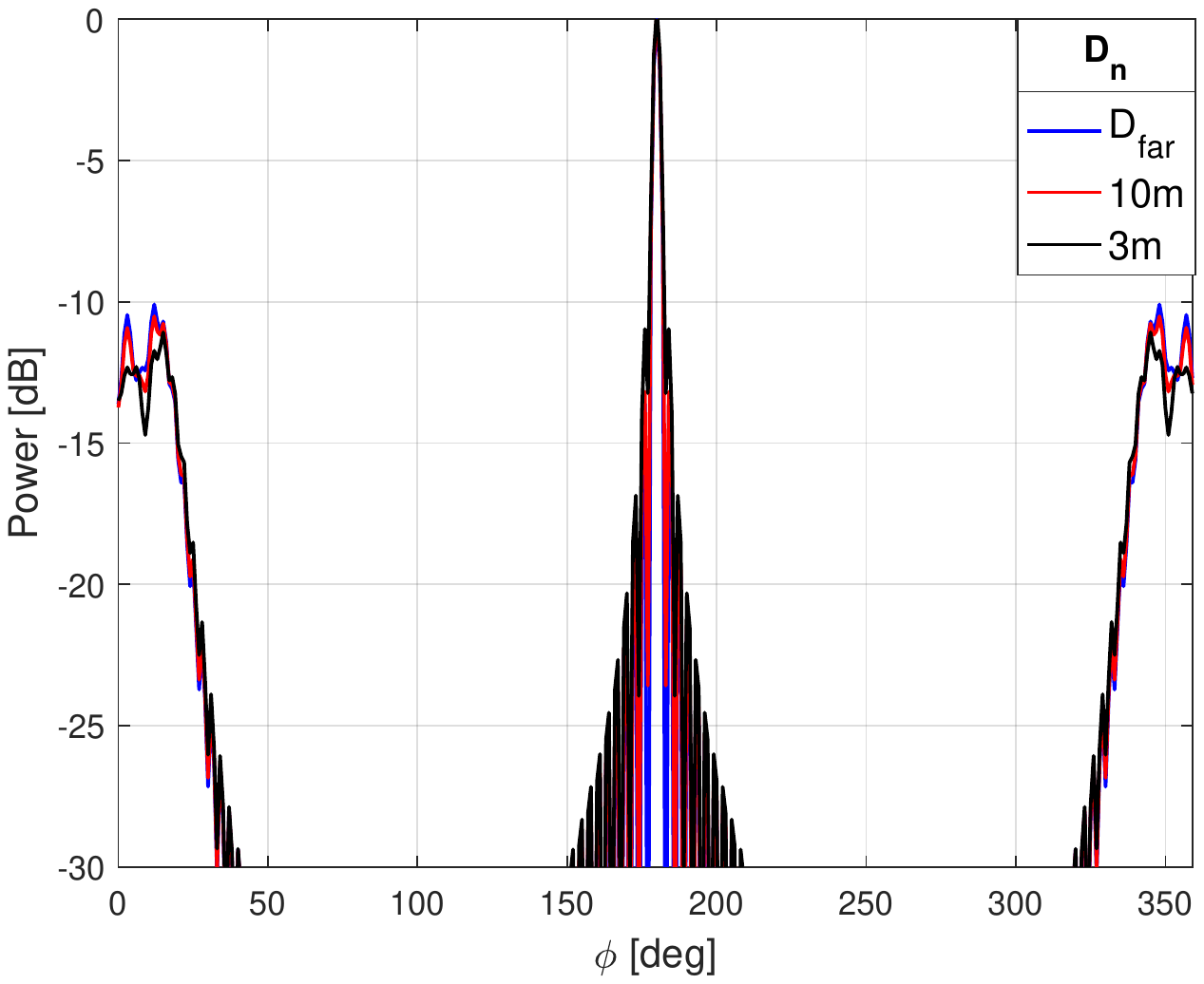}}
		\par\end{centering}
	
	\caption{The unit FIBF beam pattern $\underline{v}_{n}(f,\phi)$ in the UCA phase mode space for variant $D_n$ with $\theta_n = \pi/2$, $\phi_n = \pi$, $r$ = 0.5 m, $f$ = 29 GHz and $P=720$. \added{The calculated Fraunhofer far-field distance $D_{far} = 2(2r)^2/\lambda$ is around 193 m.}}
	\label{bp_fi_dis}
\end{figure}

 \added{Similar to the discussions about the unit beam pattern term $v_n(f,\phi)$ of CBF in section \ref{cb_ff}, we study the property of $\arrowvert \underline{v}_n(f,\phi)\arrowvert$ for various distance $D_n$. With the same simulation setting as for CBF in Fig. \ref{bp_cb_dis}, we can plot the unit FIBF beam pattern $\underline{v}_n(f,\phi)$ in Fig. \ref{bp_fi_dis}. The results show that the beam pattern is insensitive to phase error introduced by the spherical wavefront in the near-field conditions, where an approximately constant beam pattern is achieved for different distance $D_n$ within a large dynamic range.}
 
\begin{figure}
	\begin{centering}
		\textsf{\includegraphics[bb=100bp 260bp 695bp 565bp,clip,scale=0.6]{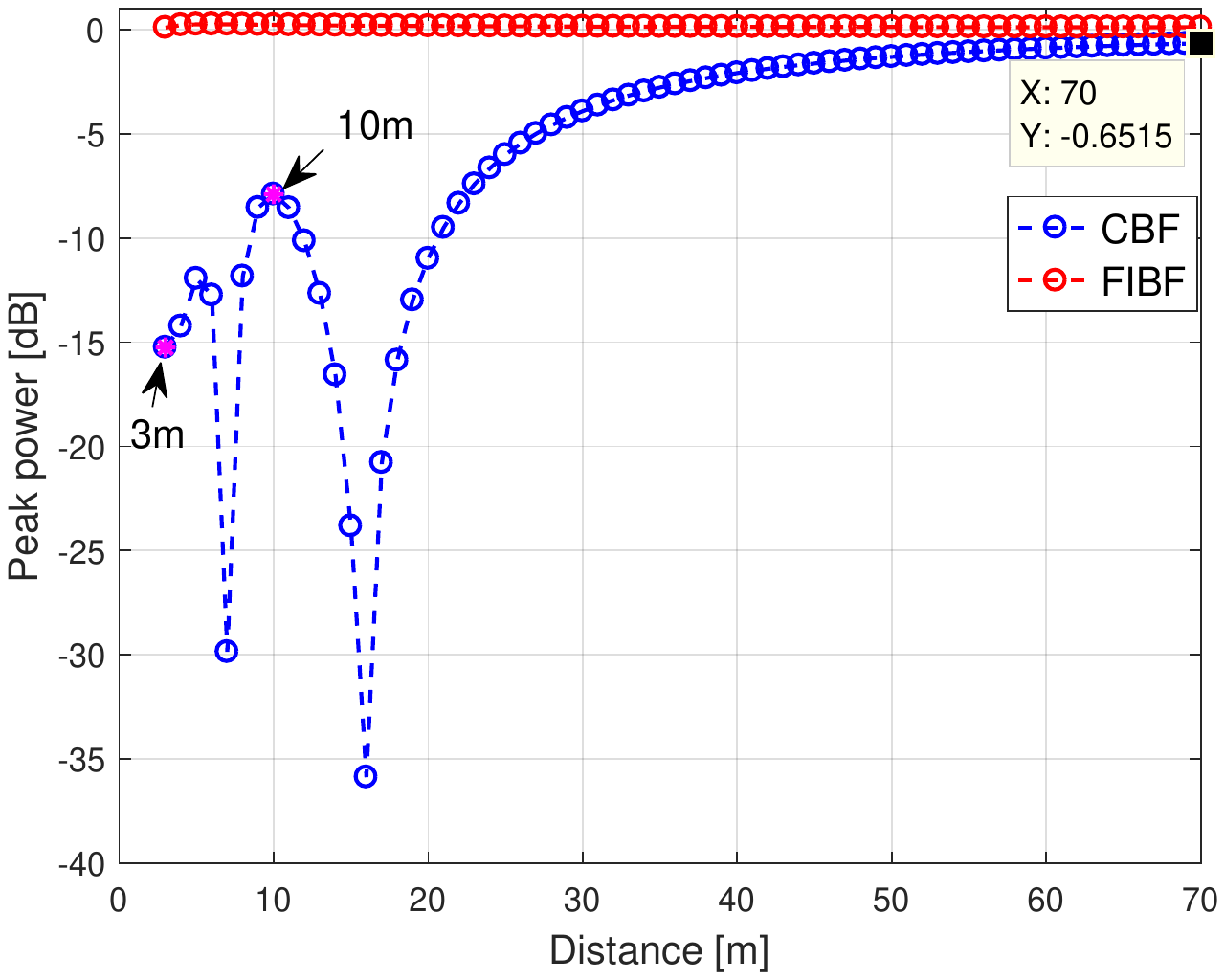}}
		\par\end{centering}
	
	\caption{The main beam peak varies with $D_n$ for CBF and FIBF, where $\theta_n = \pi/2$ and $\phi_n = \pi$, \added{$f=29$ GHz and $P=720$ are set for the UCA}.}
	\label{cbf_fibf_dis2}
\end{figure}

 \added{We further investigate the power loss in the target direction under near-field conditoins for FIBF and the results are shown by the red dotted curve in Fig. \ref{cbf_fibf_dis2}. It shows that the power values of the target direction are approximately unchanged with different distances $D_n$. } 
 
 \added{The unit FIBF beam pattern peaks in the target direction and the peak value keeps approximately constant (approximates to 1) for various distances setting, as shown in Fig. \ref{cbf_fibf_dis2}. It indicates that the path can be accurately detected with FIBF method under either near-field or far-field conditions.}

\added{As mentioned earlier, the beamformer in the UCA phase mode space with the modified compensation filter works in 3D propagation scenarios when the elevation angle is not restricted to the UCA plane. As detailed in \cite{zhang2017fi}, the beam patterns are approximately constant with different elevation angles. When the elevation angle gets away from the UCA plane, i.e. with $\varDelta \theta_n = |\theta_n - 90^o|$  getting larger, the main beam peak drops slightly and the sidelobes of the beamforming pattern at $\phi = \phi_n \pm \pi$ becomes broader. }

The simulation results indicate that the beamformer in the UCA phase mode space with the modified compensation filter $\hat{G}_m(f)$ can also be applied in near-field \added{3D} scenarios, since the beamformer pattern is insensitive to the introduced phase errors. As discussed in \cite{zhang2017fi}, the joint PADP can be directly obtained with the modified FIBF, which is simple, effective and robust. However, the resulting PADP suffers from high sidelobes,  as shown in Fig. \ref{bp_fi_dis}. In the next section, a FIBF algorithm with the successive cancellation of the detected paths is detailed, with the objective to eliminate the high sidelobes of the dominant paths.

\section{Proposed beamspace UCA with the successive cancellation scheme for near-field scenarios}

As discussed, the beamspace UCA with the modified compensation filter $\hat{G}_m(f)$ can achieve constant beam patterns, insensitive to  the distances between the scatterers and the UCA center, which makes it suitable for angle estimation in near-field scenarios. In this section, a novel  FIBF  based on successive cancellation principle to estimate azimuth angle, delay and power of each multipath component is proposed for near-field scenarios, where the paths are detected one by one with the power values in a  descending order. 

As mentioned in paper \cite{zhang2017fi}, the PADP with modified FIBF can be directly obtained by performing IFT of $\underline{B}(f,\phi)$ in (\ref{eq:aoa_fi_1}) as,
\begin{equation}
\begin{split}
\underline{b}(\tau,\phi)&= \sum_{f=f_1}^{f_L}\underline{B}(f,\phi) \exp(j2\pi f \tau)  \\          
&= \sum_{n=1}^{N}\sum_{f=f_1}^{f_L}\underline{v}_n(f,\phi) \cdot \alpha_n \exp[j2\pi f (\tau-\tau_n)]\\
&=\sum_{n=1}^{N} \underline{b}_n(\tau,\phi),
\end{split}
\label{eq:padp}
\end{equation}
where $\underline{b}_n(\tau,\phi)$ is the PADP of the $n$-th path.

In the ideal case, the PADP mimics the Dirac delta function which peaks at the unique angle-delay positions, i.e. $(\phi_n, \tau_n)$ with desired power values. However, as discussed, $\underline{v}_n(f,\phi)$ presents strong sidelobes at angle around $\phi_n \pm \pi$ for $n  \in [1,N]$. Therefore, weak desired paths might be buried by undesired sidelobes of strong paths. Below, a novel algorithm following the successive cancellation principle is proposed to tackle this problem.

\textcolor{black}{The array element response vector $\mathbf{H}(f)$ is defined as
\begin{equation}
\mathbf{H}(f) =  [H_1(f);...;H_P(f)],\label{eq:h_el_v}
\end{equation}
where $H_p(f)$ was defined in (\ref{eq:element frequency response}).}

\added{	In the following algorithm description, we add the superscript numbers to the array element response vector $\mathbf{H}(f)$ and PADP $\underline{b}(\tau,\phi)$ to indicate that these terms need to update for each iteration. The superscript number $q$ denotes that $q$ path(s) are removed. For example, $\mathbf{H}^{1}(f)$ denotes the array element response vector with 1 path removed and $\underline{b}^2(\tau,\phi)$ represents the PADP with 2 paths removed. Typically, the original array element response vector and PADP are represented with superscript number $q=0$, i.e. $\mathbf{H}^{0}(f)$ and $\underline{b}^0(\tau,\phi)$, respectively. }

The procedure is detailed as below:
\begin{itemize}
\item[1.] Based on the \added{current array element response vector $\mathbf{H}^{0}(f)$}, we apply equations (\ref{eq:hm}), (\ref{eq:Gm}), (\ref{eq:aoa_fi}) and (\ref{eq:padp}) to obtian the current PADP \added{$\underline{b}^0(\tau,\phi)$}.
\item[2. ] \added{We find the peak location in the current PADP $\underline{b}^0(\tau,\phi)$, which gives the delay and azimuth angle estimation of path 1, i.e. $\hat{\tau}_1$ and $\hat{\phi}_1$, respectively. By taking the estimated delay and azimuth angle back to the original PADP $\underline{b}(\tau,\phi)$, we can obtain the amplitude estimation of path 1, i.e. $\hat{\alpha}_1 = |\underline{b}(\hat{\tau}_1,\hat{\phi}_1)|$. } 
\item[3.] We remove \added{path 1} from the array element response vector \added{$\mathbf{H}^0(f)$} to obtain the updated array element response $\mathbf{H}^{1}(f)$ as detailed below.
\begin{itemize}
	\item[3.1.] The frequency response vector of the UCA elements corresponding to path 1, can be synthesized based on the detected path parameters under plane-wave assumption as: 
	\begin{equation}
	\hat{\mathbf{H}}(f) =  \hat{\mathbf{a}}(f,\hat{\phi}_1)\cdot \hat{\alpha}_1  \exp(-j 2 \pi f \hat{\tau}_1),\label{eq:h_el_1}
	\end{equation}
	where $\{\hat{\alpha}_1, \hat{\phi}_1, \hat{\tau}_1\}$ are parameters estimated in step 2 and $\hat{\mathbf{a}}(f,\hat{\phi}_1) \in \mathbb{C}^{P\times 1}$ is the array manifold under plane-wave condition for path 1. The $p$-th entry of $ \hat{\mathbf{a}}(f,\hat{\phi}_1)$ is given by
	\begin{equation}
	\hat{a}_{p}(f,\hat{\phi}_1) = \exp[jkr \cos(\hat{\phi}_1-\varphi_p) ].\label{eq:a_s_1}
	\end{equation}
	
	\item[3.2.] The synthetic channel impulse response (CIR) vector \added{$\hat{\mathbf{h}}(\tau)$} corresponding to \added{path 1} over array elements and the \added{current CIR vector ${\mathbf{h}}^0(\tau)$} can be directly obtained via performing IFT of \added{$\hat{\mathbf{H}}(f)$ and ${\mathbf{H}}^0(f)$}, respectively.  
	
	\item[3.3.] Generate a label vector \added{$\mathbf{s}( \tau)$} with the same size as $\hat{\mathbf{h}}(\tau)$, where the $p$-th entry is obtained as

   \begin{equation}
 s_{p}(\tau) = \left\{
	\begin{array}{ll}
	0, & \arrowvert\hat{h}_{p}(\tau)\arrowvert > \hat{\alpha}_1 \cdot 10^{\frac{-\eta_t}{20}} \\
	1, & otherwise \\	
	\end{array} \right.
	\label{eq:s}
	\end{equation}
	where \replaced{$\eta_t$} {} denotes the \replaced{threshold value}{} in decibels \added{ and $\hat{h}_{p}(\tau)$ the $p$-th entry of synthetic CIR vector $\hat{\mathbf{h}}(\tau)$}. The entries of $\hat{\mathbf{h}}(\tau)$ with dominant power values are labelled to $0$ in the label vector $\mathbf{s}(\tau)$.
	
	\added{The objective of the label vector $\mathbf{s}(\tau)$ is to mark the trajectory of the estimated path in the CIRs over array elements. As explained in step 3.1 and 3.2, we can reconstruct the path trajectory over array elements by using the estimated delay, azimuth angle and amplitude of the path under plane-wave assumption. The label vector would enable us to remove the estimated path from the CIR over array elements in Step 3.4.}
	
	\item[3.4.] \added{Remove path 1 from the current CIR vector $\mathbf{h}^0(\tau)$ to obtain the updated CIR vector $\mathbf{h}^1(\tau)$ by}
	 \begin{equation}
	 \mathbf{h}^1(\tau) = \mathbf{h}^0(\tau) \odot \mathbf{s}(\tau), \label{eq:h_el_11}
	 \end{equation}
	 where $\odot$ denotes element-wise multiplication.
	 
	 \item[3.5.] By performing Fourier transform (FT) on the updated CIR vector $\mathbf{h}^1(\tau)$, we can obtain the updated array element response vector $\mathbf{H}^{1}(f)$. 
\end{itemize} 
 \item[4.] Repeat the above steps until the estimated power of the path is not within the preset dynamic range.
\end{itemize}

\added{Note that the superscript numbers of the array element response vector and PADP increase 1 for each iteration. For instance, in the $n$-th iteration, the current array element response vector and PADP are $\mathbf{H}^{n-1}(f)$ and $\underline{b}^{n-1}(\tau,\phi)$, respectively. After step 3.5, we obtain the path parameters $\{ \hat{\alpha}_n, \hat{\phi}_n, \hat{\tau}_n\}$ and the array element response vector is updated to $\mathbf{H}^{n}(f)$. }

\added{It is also noted that in step 2, the amplitude estimation of the path is based on the original PADP $\underline{b}(\tau,\phi)$ without superscript, which is kept unchanged for all iterations. Therefore, the amplitude estimations of all the paths are obtained based on the origianl PADP instead of the updated PADP.  The reason is that, when the paths e.g. path $n_1$ and $n_2$ with $n_1>n_2$, have similar or same delays, the trajectories of the two paths are overlapped. As a result, the trajectory of path $n_2$ will be partly removed as we intend to remove path $n_1$ in step 3.4. Thus the amplitude estimation of path $n_2$ based on the updated PADP will be underestimated. This will be further illustrated in the simulation section.}

The whole procedure of the proposed path estimation algorithm is summarized in Algorithm \ref{euclid}. \added{Note that $\eta$ is the preset dynamic range in decibel.}
\begin{algorithm}
	\SetKwInOut{Input}{Input}
	\SetKwInOut{Output}{Output}
	\caption{The proposed algorithm}\label{euclid}
	\Input{$\mathbf{H}(f)$}
	\Output{$\{ \hat{\alpha}_n, \hat{\phi}_n, \hat{\tau}_n\}, n \in [1, N]$}
	\BlankLine 
	$n \coloneqq 1$\; 
	\replaced{$\hat{\alpha}_{n-1} \coloneqq 1 $ , $\hat{\alpha}_{max} \coloneqq 1 $ }{};  
		  
    $\mathbf{H}^{n-1}(f) \coloneqq \mathbf{H}(f)$, $\mathbf{h}^{n-1}(\tau) \coloneqq  IFT(\mathbf{H}(f))$\;
    	\While{ $\hat{\alpha}_{n-1} > \added{(\hat{\alpha}_{max} \cdot 10^{-\eta/20})} $}{
  		Perform phase mode space beamforming based on $\mathbf{H}^{n-1}(f)$ and obtain the PADP $\underline{b}^{n-1} (\tau,\phi)$.  // Eqs. (\ref{eq:hm}), (\ref{eq:Gm}), (\ref{eq:aoa_fi}) and (\ref{eq:padp})\; 
    	Detect the strongest path of PADP $\underline{b}^{n-1} (\tau,\phi)$ and obtain the path parameters $\{ \hat{\alpha}_n, \hat{\phi}_n, \hat{\tau}_n\}$\; 
    	Remove the detected path from CIR vector $\mathbf{h}^{n-1}(\tau)$ and update the CIR vector to $\mathbf{h}^{n}(\tau)$ \; 
    	$\hat{\alpha}_{max} \coloneqq  \hat{\alpha}_{1}$, $n \coloneqq  n + 1$. 
    }	 
\end{algorithm}	  

The basic principle of the low-cost successive cancellation scheme is that the propagation delay value among UCA array elements are insensitive to $D_n$ (i.e. distance between the array center and the $n$-th scatterer location) \added{and elevation angles}. Therefore, we can remove the detected path based on the azimuth and delay values under the plane-wave assumption. To investigate the effectiveness of the idea, \added{a single path scenario is considered here.} We can utilize the residual power rate $R_p$ \added{to evaluate how effective the detected path is removed from the CIR,} which is defined as
\begin{equation}
R_p = \frac{ \big\lVert vec{ \big\{\mathbf{h}^1(\tau)  \big\}} \big\lVert ^2}{\big\lVert vec{ \big\{\mathbf{h}(\tau) \big\}} \big\lVert ^2}\times 100 \%,\label{eq:r_p}
\end{equation}
\added{where $vec\{\cdot\}$ represents vectorization of a matrix and $\lVert \cdot \lVert $ indicates the Euclidean norm of the vector.}

\added{The high residual power rate means that the deviation between the trajectory of the synthetic CIR vector $\hat{\mathbf{h}}(\tau)$ and the true CIR vector $\mathbf{h}(\tau)$ is large, and the trajectory of the detected path will not be properly removed in the updated CIR vector $\mathbf{h}^1(\tau)$. The consequence is that we will estimate the residual trajectory as the fake path if the residual power is within the preset dynamic range.}

\added{Below, we simulate a single path case to illustrate the residual power rate $R_p$ w.r.t bandwidth, distance $D$ and the elevation angle $\theta_o$. In the simulation, a single path with a fixed incident azimuth angle $\phi_0 = 180^o$ impinging at a UCA of radius $r = 0.5$ m is set. The UCA consists of 720 elements with half-wavelength spacing. Besides, we set the center frequency to $29$ GHz, the bandwidth from $400$ MHz to $3$ GHz, distance $D$ from 3 m to 30 m and elevation angle from $90^o$ to $120^o$. }

\begin{figure}
 	\begin{centering}
 		\textsf{\includegraphics[bb=105bp 275bp 477bp 568bp,clip,scale=0.69]{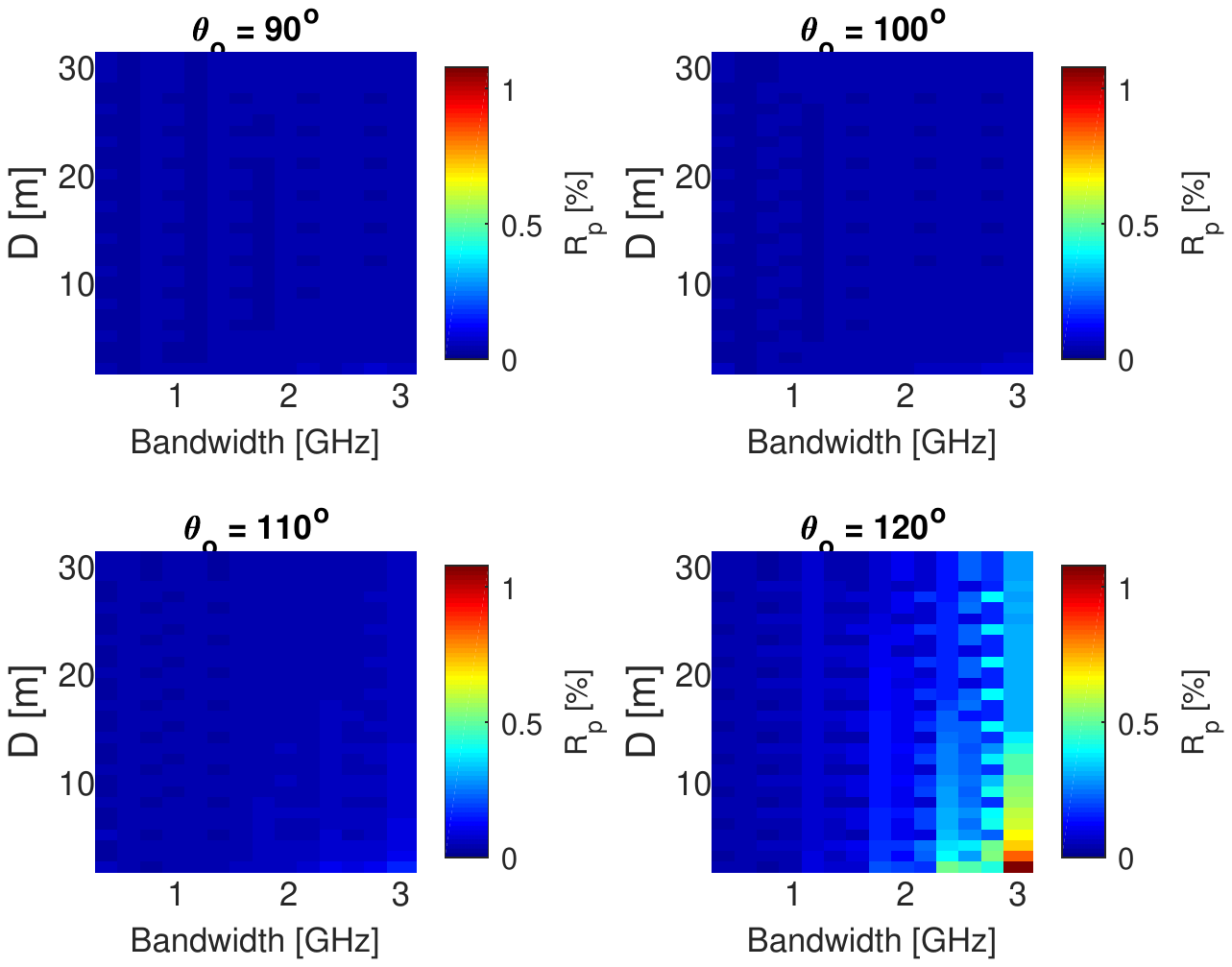}}
 		\par\end{centering}	
 	\caption{\added{The residual power rate $R_p$ varies with bandwidth, distance $D$ and elevation angle $\theta_o$, where $f_c = 29$ GHz, $\phi_o = \pi$ and $\tau_o = 0$ ns.}}
 	\label{b_dis}
 \end{figure}

\added{The impact of system bandwidth, distance $D$ and elevation angle $\theta_o$ on the residual power rate is shown in Fig. \ref{b_dis}.  }
\textcolor{black}{\begin{itemize}
	\item For a given bandwidth and elevation angle $\theta_o$, $R_p$ decreases as $D$ increases due to the fact that the larger the $D$ is, the closer we approximate the plane-wave condition, resulting in a smaller $R_p $.
	\item For a given distance $D$ and elevation angle $\theta_o$, $R_p$ increases as bandwidth broadens. The wider the bandwidth, the higher the delay resolution we have. As a result, the larger reconstructed CIR error will be detected as the bandwidth becomes wider, which leads to a larger $R_p$.
	\item For a given bandwidth and distance $D$, $R_p$ increases when elevation angle $\theta_o$ gets further away from the UCA plane, i.e. with $\varDelta \theta =\arrowvert\theta_o -90^o \arrowvert $ becoming larger. This is due to the fact that the synthetic CIR was calculated with $\hat{\theta}_o = 90^o$ in (\ref{eq:h_el_1}) and (\ref{eq:a_s_1}). The larger the elevation angle away from the UCA plane is, the larger error will be introduced to the reconstructed CIR. The maximum $R_p$ are $0.06\%$, $0.07\%$, $0.16\%$ and $1.1\%$ when elevation angle $\theta_o$ are set to $90^o$, $100^o$, $110^o$ and $120^o$, respectively.
\end{itemize}
}

\added{For a practical measurement setting with bandwidth less than
	$3$ GHz, the scatterer distance larger than 3 m and the elevation angle $\arrowvert\theta_o -90^o \arrowvert \leqslant 30 ^o$, the residual power rate is up to $1.1 \%$, as shown in Fig. \ref{b_dis}. Therefore, the proposed algorithm works well for practical measurement settings. The proposed cancellation scheme is low-cost, since it only requires the estimated delay and azimuth angle values to remove the detected path. It is effective and robust, as demonstrated in the numerical simulations. }

 \added{As a summary, the basic principle of the low-cost successive cancellation scheme is that the propagation delays among the UCA elements are insensitive to the elevation angle of the path and the distance between the UCA center and the scatterer location. Therefore, we can effectively remove the detected path based on azimuth angle and delay of the path in the updated delay profile among elements. Due to the sparsity of mmWave channels, typically only a few iterations are needed to extract all multipath components. As shown, for each iteration, we obtain the delay and azimuth angle estimate based on the power spectra, which essentially is calculated from the one-dimensional beamforming operation in (\ref{eq:aoa_fi}) and IFT operation in (\ref{eq:padp}). Therefore, the computational cost is significantly lower, compared to typical high resolution algorithms, where expensive joint search in multiple parameter domains is required.}
 
\section{Simulation results} \label{simulation}
\added{In the simulation, we consider a UCA composed of 720 isotropic antennas with radius $0.5$ m and half-wavelength element spacing. The frequency band is from $28$ - $30$ GHz with $750$ frequency points. We simulate a representative yet critical scenario for the channel estimation algorithm, where path 1 and path 3 have the same impinging azimuth angle $\phi$ at $90^o$. Path 2 has an incident angle of $270^o$, yet it has the same delay as path 3. In addition, the elevation angles are not strictly confined to the UCA plane as detailed in Table \ref{tab:table1}.}
The critical scenario is intentionally set to demonstrate the robustness of the algorithm.

\begin{figure}
	\begin{centering}
		\textsf{\includegraphics[bb=95bp 225bp 500bp 630bp,clip,scale=0.65]{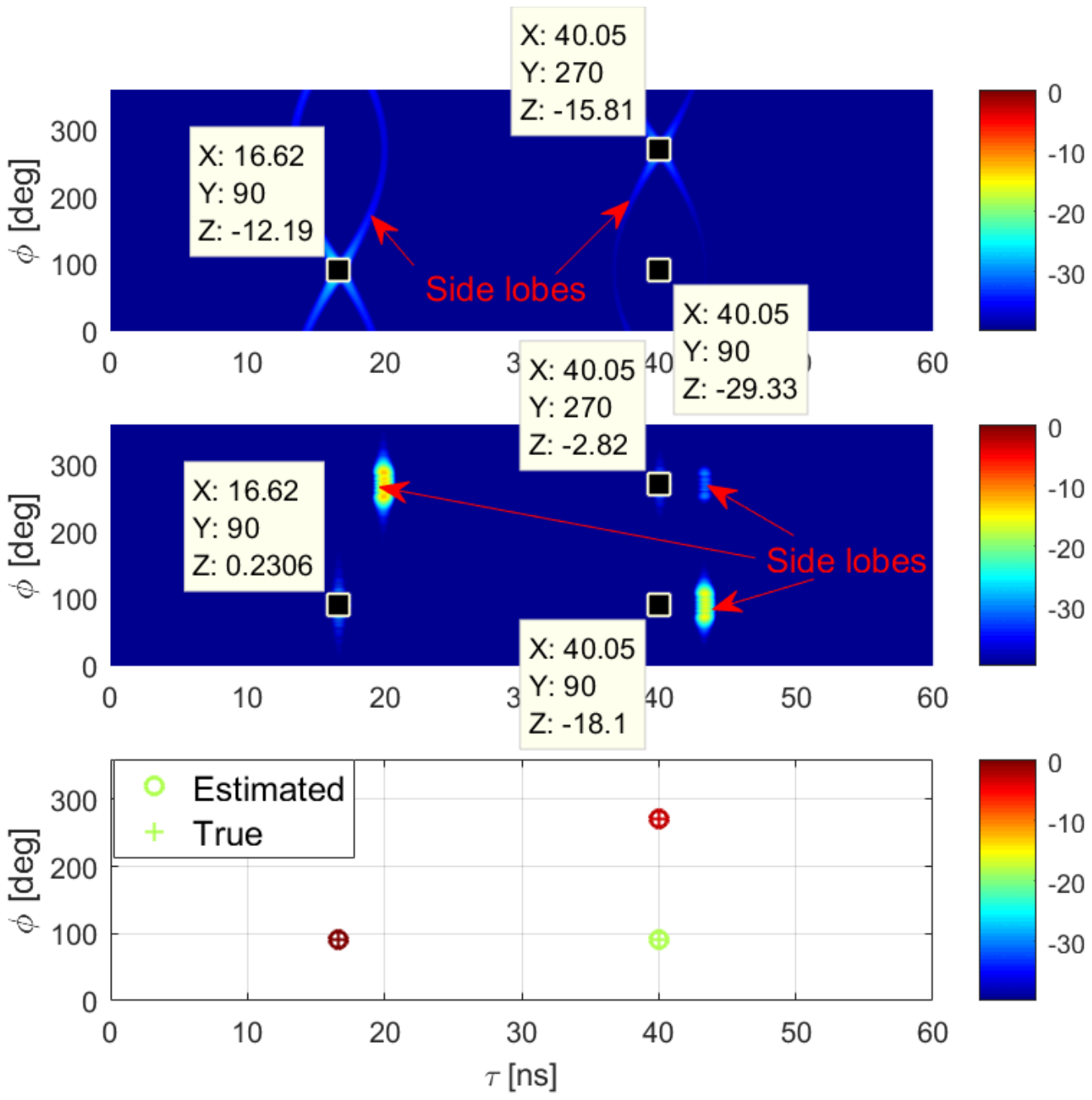}}
		\par\end{centering}
	\caption{\added{The PADPs with CBF (top), FIBF (middle) and the proposed successive FIBF (bottom).}}
	\label{ad_sim}
\end{figure}

\replaced{Three beamforming algorithms, i.e. CBF \cite{fan2016eurasip}, FIBF \cite{zhang2017fi} and the proposed algorithm, are compared in Fig. \ref{ad_sim}.}{} As shown in Fig. \ref{ad_sim} (top), though target paths can be \added{roughly} detected in the PADP, the CBF algorithm under far-field assumption presents two major drawbacks as explained in Section \ref{cb_ff}, i.e. susceptible to strong joint side lobes and main lobe distortions. The power loss of the main lobes up to around \replaced{$13$} {} dB and the concave main lobe (path 1) can be observed in the figure due to the far-field CBF applied in the near-field scenario. The detailed of the concave main lobe can be observed in Fig. \ref{bp_cb_dis} for $D_n = 3$m. Note that the joint side lobes for the weak path (i.e. path 3) are not shown in the figure due to the limited dynamic range ($40$ dB) set in the simulation. The PADP with the FIBF algorithm is shown in Fig. \ref{ad_sim} (middle). It shows that the main lobes are not distorted and the peak values are accurate as explained in Fig. \ref{bp_fi_dis}. However, as discussed, the FIBF suffers from strong side lobes around azimuth angle $\phi_n \pm \pi$ for $n \in [1,N]$ with a shifted delay. The estimated parameters $\{\hat{\alpha}_n, \hat{\phi}_n, \hat{\tau}_n\}, n \in [1,N]$ with the proposed algorithm are shown in Fig. \ref{ad_sim} (bottom). The estimated parameters agree well with the targets, with a deviation in path power within $0.3$ dB. The small power deviation is caused by the power variance over the UCA elements in the near-field scenario. 
\added{Based on the Friis free space propagation equation, the maximal power deviation over the UCA elements for a single path case can be evaluated by the ratio of the maximal and minimal power across the UCA elements, i.e., $\rho = \frac{D+r}{D-r}$,  where $r$ denotes the radius of the UCA and $D$ the distance between the scatterer and the UCA center with $D>r$. In the far-field case (i.e. $D\gg r$), we have $\rho \approx 1 $. However, the power over the UCA elements varies in the near-field case. }

Below, we detail the procedure how the path parameters are estimated with the proposed algorithm.  

\begin{figure}
	\begin{centering}
		\textsf{\includegraphics[bb=80bp 180bp 500bp 665bp,clip,scale=0.58]{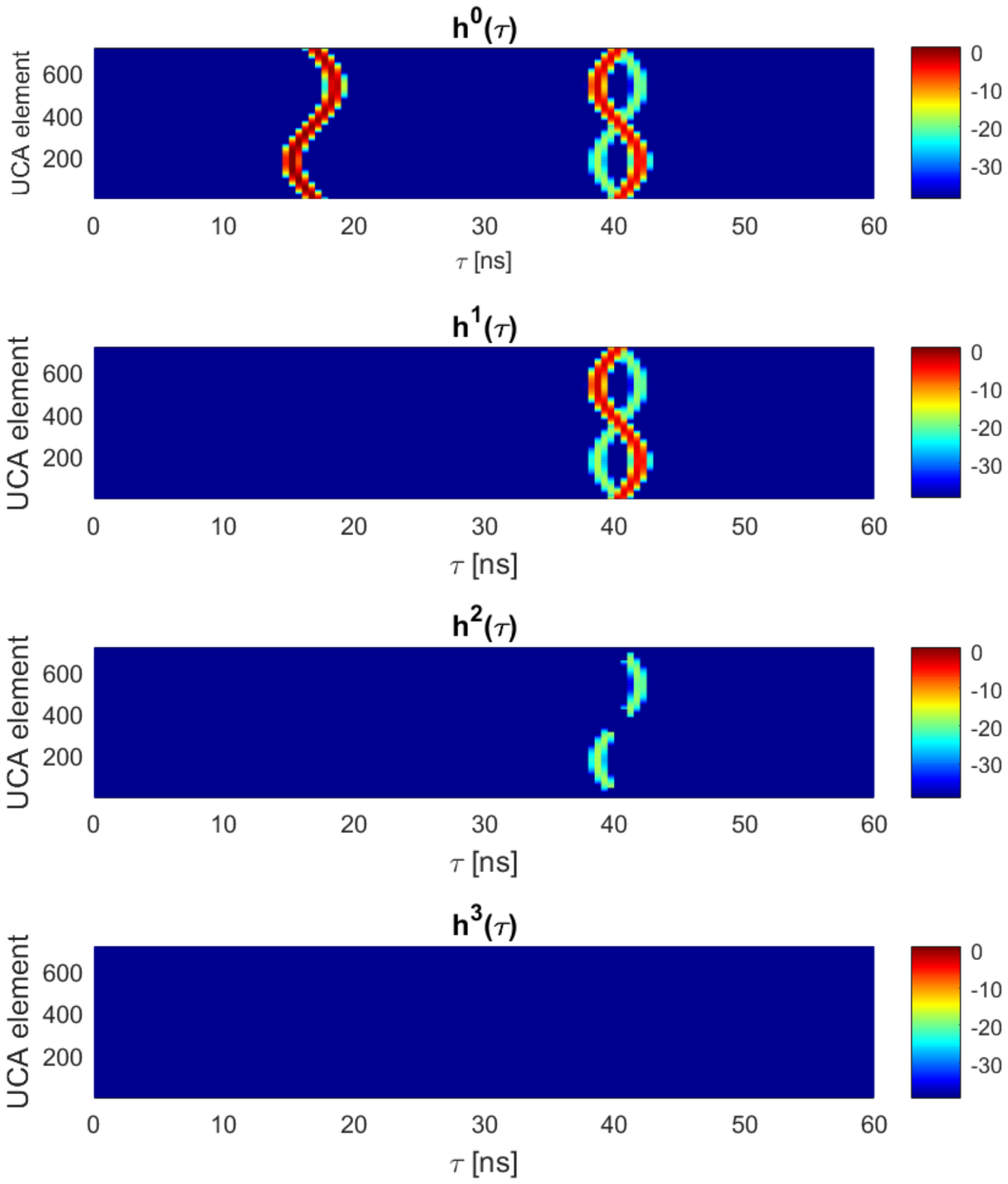}}
		\par\end{centering}	
	\caption{\added{The CIR vectors of $\mathbf{h}^{n-1}(\tau), n \in [1,4]$, where the superscript denotes the $(n-1)$ path(s) are removed.}}
	\label{el_delay}
\end{figure}

\begin{figure}
	\begin{centering}
		\textsf{\includegraphics[bb=70bp 185bp 530bp 675bp,clip,scale=0.58]{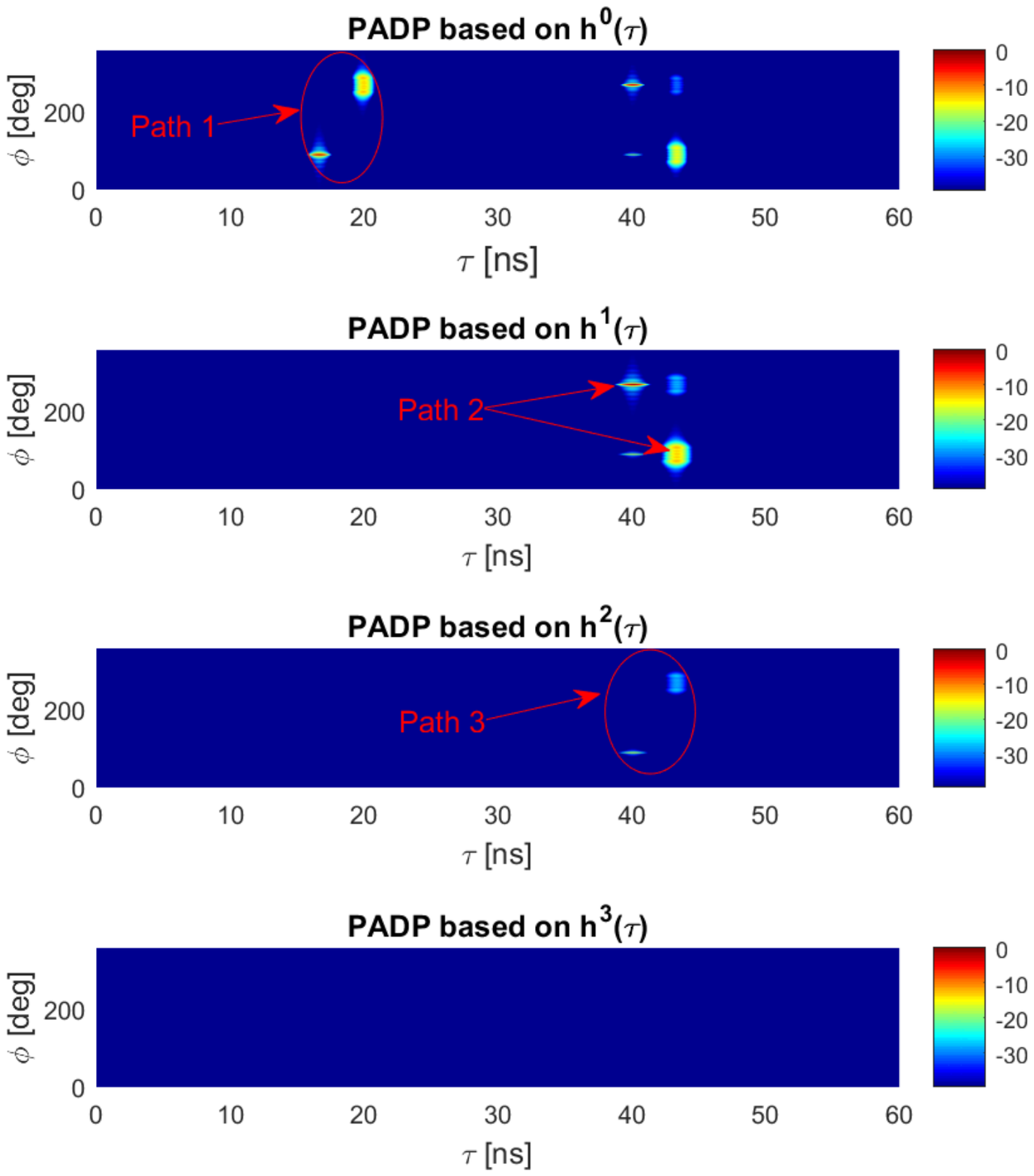}}
		\par\end{centering}	
	\caption{\added{The PADPs based on $\mathbf{h}^{n-1}(\tau), n \in [1,4]$, where the superscript denotes the $(n-1)$ path(s) are removed.}}
	\label{bp_fi}
\end{figure}

To detect the most dominant path, we can follow the procedure below: 
	\begin{itemize}
		\item[1.] We can perform the phase mode beamforming based on $\mathbf{h}^{0}(\tau)$ (i.e. the raw CIR vector $\mathbf{h}(\tau)$), as shown in Fig. \ref{el_delay} (top), and the obtained PADP $\underline{b}^0 (\tau, \phi)$ is shown in Fig. \ref{bp_fi} (top).
		\item[2.] From the PADP, we can detect the strongest path, i.e. path 1 as shown in Fig. \ref{bp_fi} (top) with path parameters $\hat{\alpha}_1 = 0.2$ dB, $\hat{\phi}_1 = 90^o$ and $\hat{\tau}_1 = 16.6$ ns.
		\item[3.] Then, we remove path 1 from the original CIR vector $\mathbf{h}^0(\tau)$ and obtain the updated CIR vector $\mathbf{h}^1(\tau)$ as shown in Fig. \ref{el_delay} \added{(upper-middle)}.
		\item[4.] In the end,  $\hat{\alpha}_{max} = \hat{\alpha}_1$ is set.
	\end{itemize} 
\added{We can repeat the above procedure to detect the second and third paths. The detected path parameters of path 2 and path 3 are $ \{\hat{\alpha}_2 = -2.8$ dB,  $\hat{\phi}_2 = 270^o,\hat{\tau}_2 = 40.1 $ ns $\}$ and $ \{\hat{\alpha}_3 = -18.1$ dB,  $\hat{\phi}_3 = 90^o,\hat{\tau}_3 = 40.1$ ns $\}$, respectively.}
In the end, a path with power value within $40$ dB dynamic range can not be found based on $\mathbf{h}^{3}(\tau)$ and therefore the channel estimation procedure is complete.

\added{We can clearly see that the detected path is removed in the updated CIR, e.g. path 1 is removed in the updated CIR vector $\mathbf{h}^{1}(\tau)$ as shown in \ref{el_delay} (upper-middle). Thus the influnce of the path is also eliminated in the updated PADP $\underline{b}^1 (\tau, \phi)$ as shown in \ref{bp_fi} (upper-middle). In this simulation, path 2 and 3 have the same delays and therefore the trajectories of the two paths are overlapped, as shown in Fig. \ref{el_delay} (upper-middle). When we remove path 2, the trajectory of path 3 will be partly removed, as illustrated in Fig. \ref{el_delay} (lower-middle). As discussed in section III, to avoid underestimating the amplitude of path 3 $\hat{\alpha}_3$, we estimate the amplitude based on the original PADP $\underline{b}(\tau,\phi)$ as shown in Fig. \ref{ad_sim} (middle).}

\begin{table}
	\caption{Path parameters}
	\label{tab:table1}
	\begin{centering}
		\begin{tabular}	{|c|c|c|c|}		
			\hline 
			Path & $1$  & $2$  & $3$  \\ 
			\hline 
			$\alpha$ [dB] & 0 & \added{-3}   & \added{-18} \\ 
			\hline 
			$\phi$ [deg] &90  &270   &90\\ 
			\hline 	
			$\theta$ [deg] &90  &95  &100 \\ 
			\hline	
			$D$ [m] &4.98   &\added{12}   &12\\ 
			\hline	
			$\tau$  [ns] &16.6  &\added{40.0} &40.0   \\ 
			\hline
		\end{tabular} 
		\par\end{centering}
\end{table}

\section{Measurement Results}  \label{measurement}

\subsection{Introduction}

To verify how well the proposed algorithm works in practice, we need
to validate it with practical measurements. The detailed description
of the measurement campaign was given in \cite{fan2016eurasip} and only
outlined here. The measurements were conducted in a typical indoor
basement \added{with the floor dimensions of $7.7$ m $\times$ $7.9$ m. The basement was empty with few objects including a metallic heater and a metallic ladder leaned against the wall.}
Both line-of-sight (LOS) and obstructed LOS (OLOS)
scenarios were considered. The OLOS scenario was created by placing
a $1.2$ m $\times$ $1.2 $ m metallic blackboard to block
paths in LOS directions.

\added{A wideband biconical antenna with a gain of $6$ dB at 28-30 GHz was used at
	the transmit (Tx) side. The Tx antenna was mounted $0.84$ m above the floor. While an identical biconical antenna was exploited at the receive (Rx) side.
	The Rx antenna was mounted on a turntable with the same height as the Tx antenna. The distance between the Rx antenna and the rotation center was adjusted to $0.5$ m. Then a virtual UCA of radius $r=0.5$ m at Rx side were obtained by rotating the Rx antenna on the turntable. The frequency response of the $p$-th UCA element was measured when the Rx antenna was rotated to the angular position $\psi_p = 2\pi\cdot(p-1)/P, p\in [1,P]$ with $P = 720$. 
	For each virtual UCA element, the frequency
	response was measured with a vector network analyzer (VNA) from 28-30
	GHz with 750 frequency points. }

\added{The rotational horn antenna
	measurements were used as a reference. In the rotational horn antenna measurements, the biconical antenna at the Rx side was replaced by a horn antenna with a gain of $19$ dB at 28-30 GHz. The horn antenna was positioned	at the rotation center of the turntable (i.e. $r=0$ m) with the same height as the Tx antenna (i.e. $0.84$ m above the floor). The same measurement settings (the same Tx antenna, the same frequency sweep and orientation sweep) adopted in biconical antenna UCA measurements were used in rotational horn antenna measurements for comparison purpose. }

Note that mutual coupling effect is not present with the virtual array measurement, which is desirable
for the  channel characterization purpose. The distance between the Tx
and the center of the Rx array is around 5 m, while the far-field
distance at 30 GHz for the UCA is around 200 m. The system bandwidth
in the measurement is larger than 500 MHz and the UCA array aperture
(i.e. 1m) is  much larger than the delay resolution multiplied by
the speed of light (i.e. 0.15 m). Therefore, for the measurement data,
both the far-field assumption and narrowband assumption are violated.
Note that \added{the antenna gains of both Tx and Rx antennas are de-embedded in measured CIRs. Further, to focus on the specular and dominant multipath components detection,} a dynamic range of 30 dB is set in the measurement section. 

\subsection{Measured results}

\begin{figure}
	\begin{centering}
		\textsf{\includegraphics[bb=100bp 270bp 480bp 567bp,clip,scale=0.58]{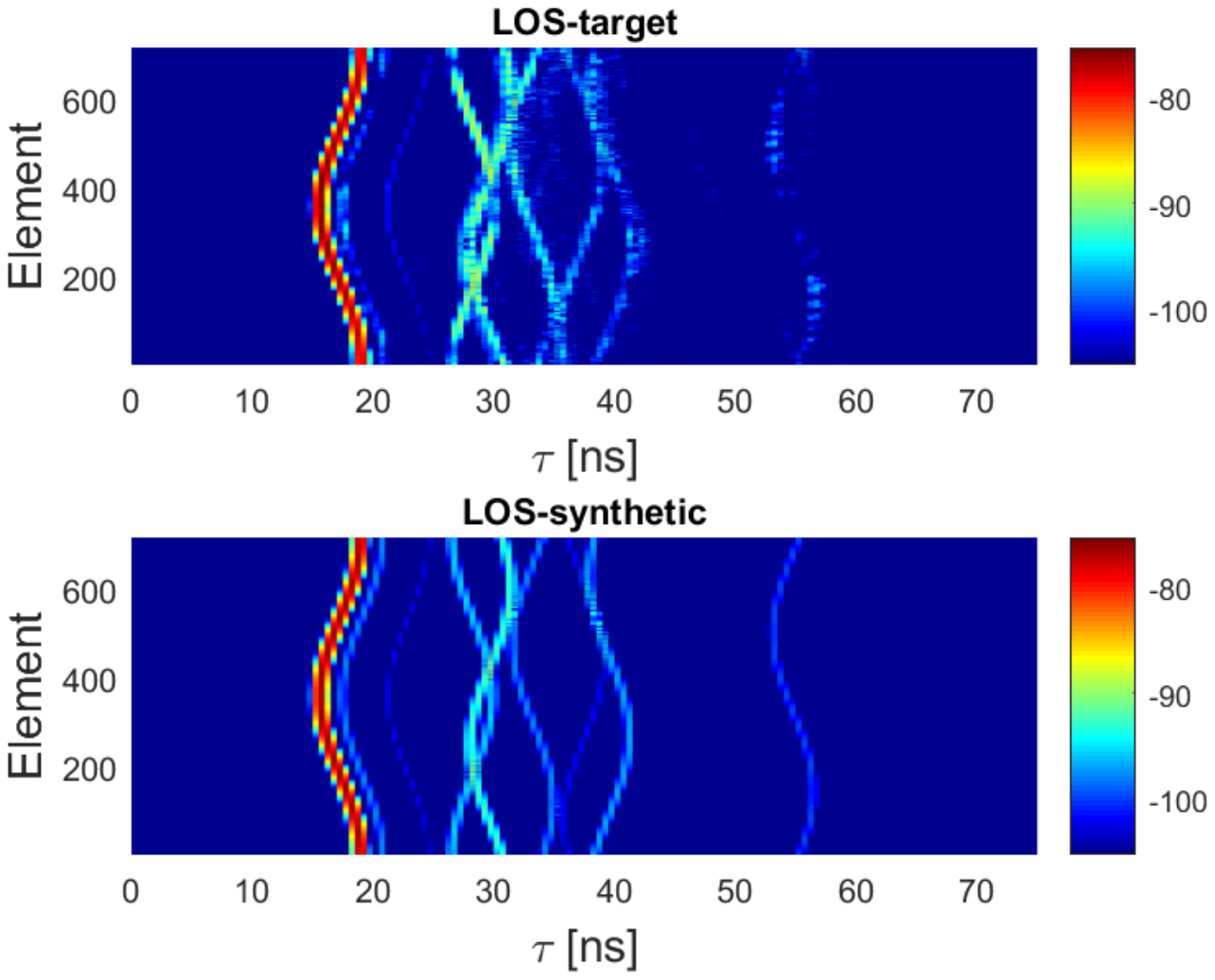}}
		\par\end{centering}
	\caption{\added{Measured CIRs (top) and synthetic CIRs (below) over virtual UCA elements
		for the LOS scenario. }}
	\label{LOS_synthetic}
\end{figure}

\begin{figure}
	\begin{centering}
		\textsf{\includegraphics[bb=100bp 270bp 480bp 550bp,clip,scale=0.58]{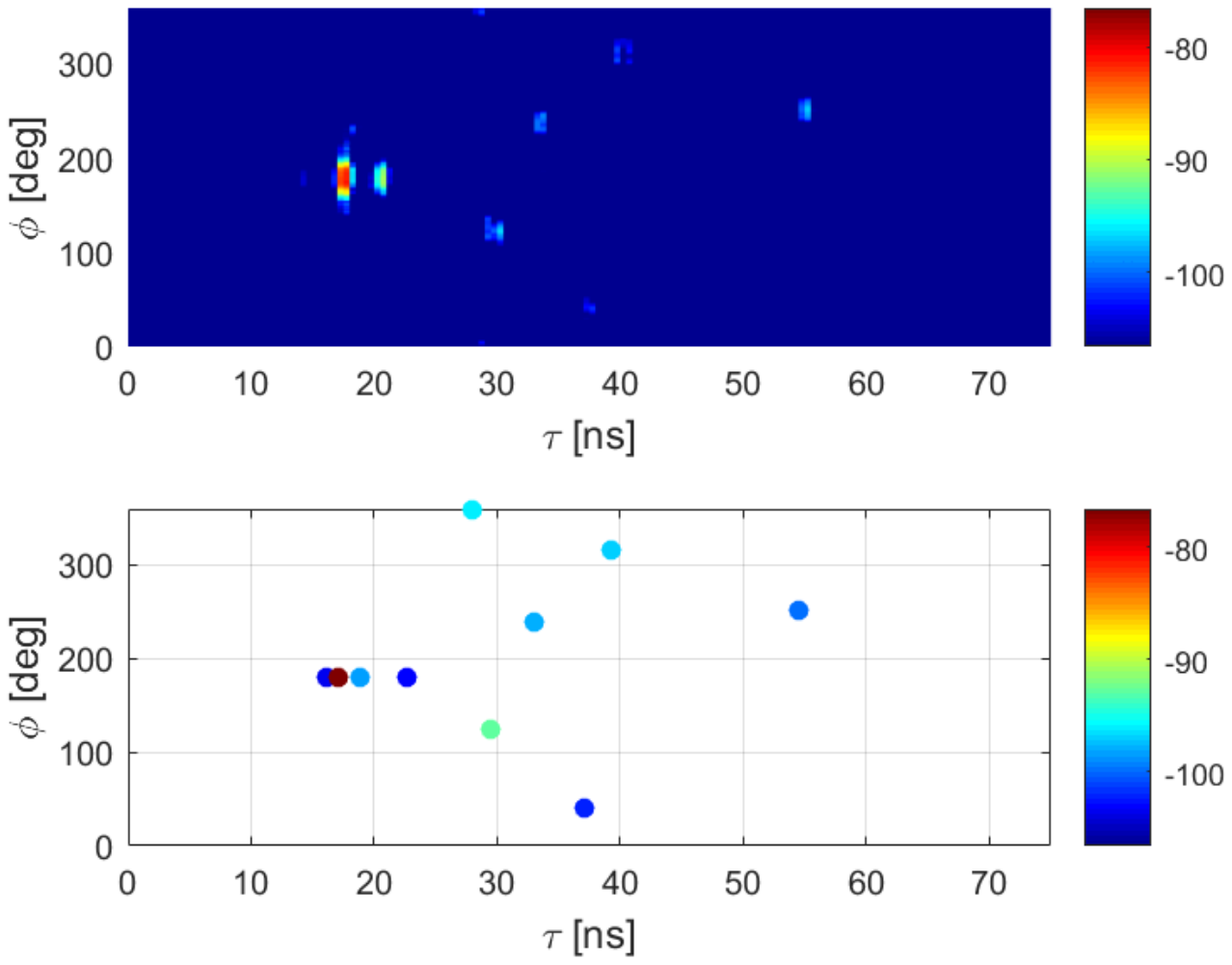}}
		\par\end{centering}
	\caption{\added{Measured PADPs with rotational horn antenna (top)
		and virtual UCA with the proposed algorithm (below) for the LOS scenario. }}
	\label{LOS_horn}
\end{figure}

The measured CIRs over virtual UCA elements (i.e. measured locations) for the LOS scenario are shown
in Fig. \ref{LOS_synthetic} (top), where a few specular paths can
be clearly detected besides the dominant LOS path. \added{The mmWave channels are more sparse compared to sub-6GHz channels. In our measurements, the measurement was performed in an empty indoor basement, with no furniture, which also results in sparse channel profiles.} The synthetic CIRs
over UCA elements  for the LOS scenario are shown in Fig. \ref{LOS_synthetic}
(below). The synthetic CIRs over virtual UCA elements are reconstructed based on detected multipath component parameters  $\{\hat{\alpha}_{n},\hat{\phi}_{n},\hat{\tau}_{n}\}$
for $n\in[1,N]$ under the plane-wave assumption. The synthetic results agree well with the measured data,
indicating a consistent estimation result. The \added{trajectories of the paths} over UCA elements match well with the measured ones, even for weak multipath components. The measured PADP
with rotational horn antenna is shown in Fig. \ref{LOS_horn} (top).
As shown in the measured results, paths having the same impinging angle
yet different delays exist due to the path  bouncing in the LOS direction.
The estimated PADP based on the virtual UCA with
the proposed algorithm for the LOS scenario is shown in Fig.
\ref{LOS_horn} (below), \added{where in total 10 paths are detected}. An excellent match of the measured PADPs between the rotational
horn antenna and virtual UCA in terms of the number of propagation paths,
azimuth angle, delay and power of each path can be observed. However, the
measured results based on rotational horn antenna suffer from wide
antenna beam-width, as expected. The proposed algorithm presents consistent
parameter estimation, with high resolution in the angle and delay
domains. 
\added{ Note that the antenna gains of the horn antenna and biconical antenna are calibrated out in the power spectra. Thus within the same power range, the same number of paths and approximately same path parameters can be observed in the plots. Furthermore, the estimated power of the LOS path (strongest path) is $-76.5 $ dB, which matches well with the calculated path loss according to Friis equation with $D = 5$ m and $f_c = 29$ GHz, i.e. $-75.7$ dB. The deviation might be introduced by the inaccureate data in antenna gains and measurement uncertainties.}

\begin{figure}
	\begin{centering}
		\textsf{\includegraphics[bb=100bp 270bp 480bp 567bp,clip,scale=0.58]{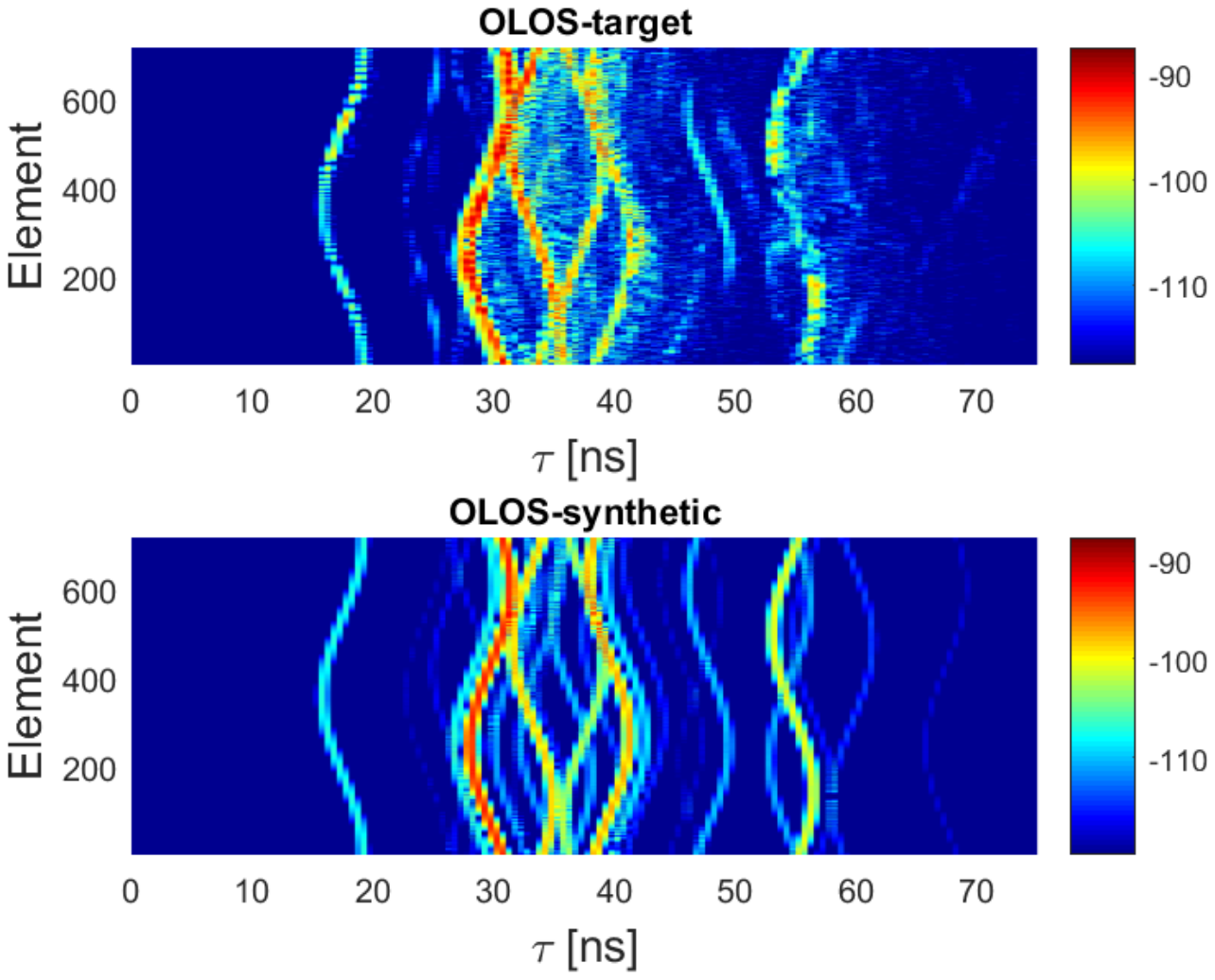}}
		\par\end{centering}
	\caption{\added{Measured CIRs (top) and synthetic CIRs (below) over virtual UCA elements
			for the OLOS scenario. } }
	\label{OLOS_synthetic}
\end{figure}

\begin{figure}
	\begin{centering}
		\textsf{\includegraphics[bb=100bp 270bp 480bp 550bp,clip,scale=0.58]{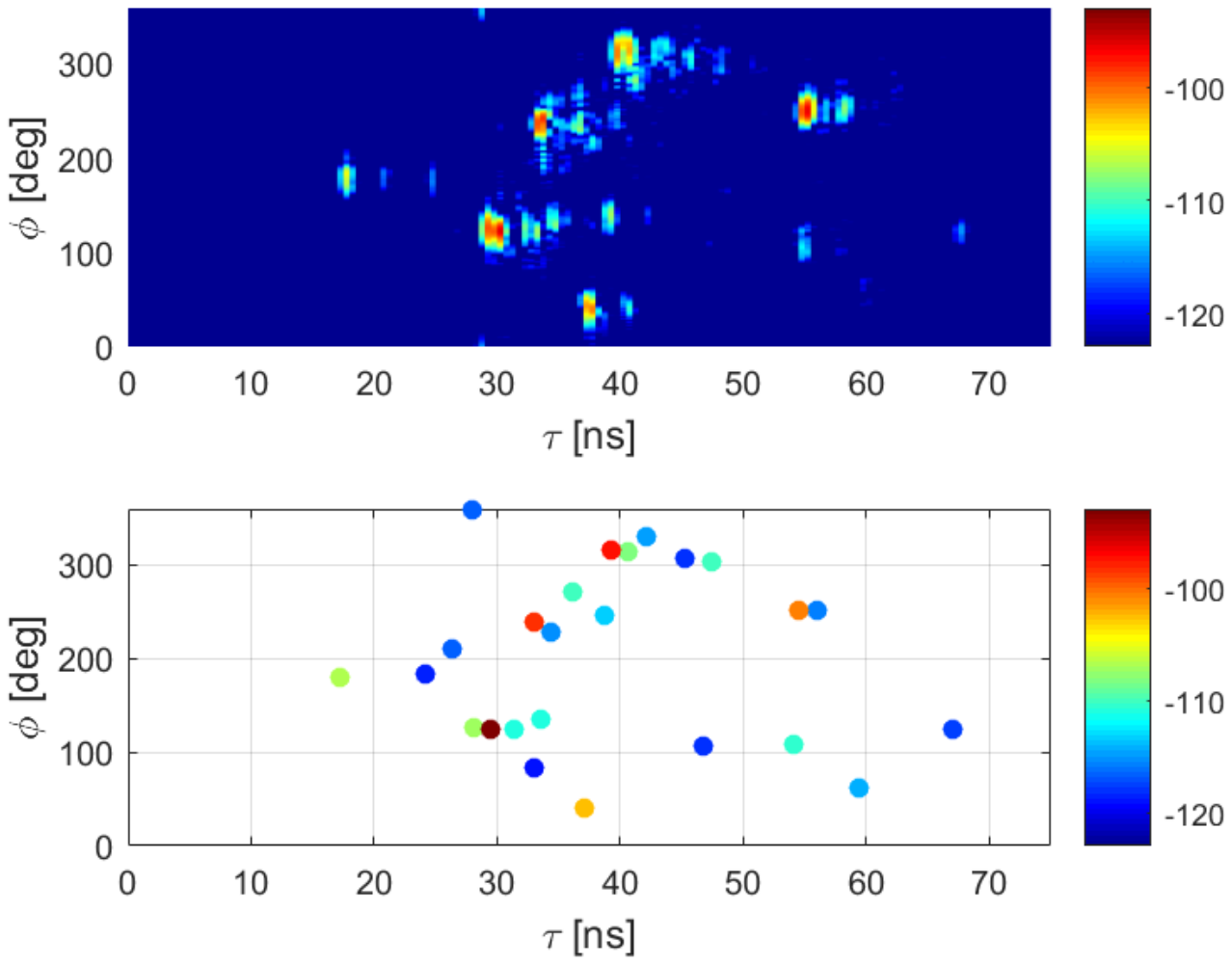}}
		\par\end{centering}
	\caption{\added{Measured PADPs with rotational horn antenna (top)
		and virtual UCA with the proposed algorithm (below) for the OLOS scenario.} }
	\label{OLOS_horn}
\end{figure}

The measured CIRs and synthetic CIRs over virtual UCA elements for the
OLOS scenario are shown in Fig. \ref{OLOS_synthetic}. A good agreement
between the synthetic and measured CIRs can still be observed,
though \added{there exists many weak multipath components in }the measured CIRs. The measured PADPs with rotational horn
antenna and virtual UCA with the proposed algorithm for the OLOS scenario
are shown in Fig. \ref{OLOS_horn}, \added{where $27$ paths in total are detected within the dynamic range of $30$ dB. Within the same power range, the channel parameters, e.g. the number of paths and path paramters, agree well between the rotational horn and the virtual UCA for
	the more critical OLOS scenario. }

\added{As observed in the measured CIRs over array elements, channel non-stationarity exists where different channel profiles can be observed by different array elements. This can be caused by several reasons, e.g. the power variation over the UCA elements in the near-field scenarios as explained; the coherent summation of unresolved multipath components due to limited system bandwidth; limited angle of view from the near-field scatterers and the measurement system non-idealities. The channel non-stationarity is not addressed in the proposed algorithm, as seen in the synthetic CIRs over UCA elements. Though channel non-stationarity has been considered in channel modeling works, e.g. \cite{he2018mobility}, it has not been considered in the channel estimation in the literature so far due to high computation complexity.  }

\section{conclusion}
Accurate knowledge of the radio propagation parameters is important for system design, applications and performance evaluation of the 5G systems. However, multipath parameter estimation  for UWB large scale antenna systems are challenging,  due to the fact that the well adopted plane-wave  and narrowband assumptions might not hold. In this paper, a novel beamspace UCA algorithm based on phase mode excitation principle is proposed. The proposed beamformer can maintain approximately same beam patterns, independent of distance between the array and scatterer location, and of the system bandwidth, making it suitable for UWB near-field scenarios. The proposed algorithm has low computational cost since it avoids expensive joint estimation in multiple parameter domains. To remove the strong sidelobes of the proposed beamspace beamformer, a novel algorithm based on the successive cancellation principle is proposed. The path cancellation is based on the fact that propagation delays among array elements are insensitive to system bandwidth and near-field effects \added{for a given elevation angle range, i.e. $\varDelta \theta = \arrowvert \theta - 90^o\arrowvert \leq 30^o$}. The cancellation scheme is effective and robust. For example, the residual power rate of less than 0.2\% can be achieved for the UCA with radius $r = 0.5$m, system bandwidth 2 GHz and measurement range  $D\geqslant 3$ m.  To demonstrate the proposed algorithm, a critical scenario is selected in the numerical simulation \added{with frequency band set the same as in the measurement campaign, i.e. 28-30 GHz}, and the results showed that all the paths can be accurately detected with less than 0.3 dB deviations for the path powers. To validate the algorithm, we applied the proposed algorithm in the practical virtual UCA measurement data in both LOS and OLOS scenarios. The synthetic CIRs obtained based on the detected multipath parameters matched well with the measured CIRs for both measurement scenarios. The detected parameters were further validated against the horn antenna reference measurements. As a summary, both numerical simulations and experimental measurements demonstrated the effectiveness and robustness of the proposed algorithm. The proposed algorithm is a general low-cost channel estimator, since it works in both near-field scenario and UWB system.

\added{Further, due to the fact that the proposed beamformer pattern is not sensitive
	to the elevation angle and scatterer location, the proposed
	algorithm would fail to detect the elevation angle and scatter
	location. A high resolution propagation parameter estimation
	algorithm, which is capable of estimating all
	propagation parameters in the 3D near-field conditions is missing
	in the literature due to the fatal computation complexity.
	Our proposed algorithm, which offers high resolution azimuth
	angle, delay and complex amplitude estimation in a low-cost
	manner, can be utilized in the initial stage of the full parameter
	estimation algorithm, e.g. the maximal likelihood estimator,
	which can significantly reduce the computation complexity
	due to the reduced searching space. This work will be carried out in a future work. }

\bibliographystyle{IEEEtran}
\addcontentsline{toc}{section}{\refname}\bibliography{Wei_library}

\end{document}